\documentclass[12pt]{article}
\usepackage{graphicx}
\usepackage{amssymb,amsfonts,amsmath}

\def\lsim{\mathrel{\rlap{\lower3pt\hbox{\hskip1pt$\sim$}}
     \raise1pt\hbox{$<$}}} 
\def\gsim{\mathrel{\rlap{\lower3pt\hbox{\hskip1pt$\sim$}}
     \raise1pt\hbox{$>$}}} 
\newcommand{\beq}{\begin{equation}}
\newcommand{\eeq}{\end{equation}}
\newcommand{\bea}{\begin{eqnarray}}
\newcommand{\eea}{\end{eqnarray}}

\begin{document}
\title{Shear Viscosity of a strongly interacting system: Green-Kubo versus Chapman-Enskog and Relaxation Time Approximations}

\author{S. Plumari$^{a,b}$, A. Puglisi$^{b}$, F. Scardina$^{b}$ and V. Greco$^{a,b}$\\
$^a$ \small{\it Department of Physics and Astronomy, University of Catania,}\\ 
\small{\it Via S. Sofia 64, I-95125 Catania (Italy)}\\
$^b$ \small{\it Laboratorio Nazionale del Sud, INFN-LNS, Via S. Sofia 63, I-95125 Catania (Italy)}\\}
\maketitle

\maketitle

\begin{abstract}

The shear viscosity $\eta$ has been calculated by using the Green-Kubo relation in the framework of a partonic transport
approach solved at cascade level.
We compare the numerical results for $\eta$ obtained from the Green-Kubo correlator with the analytical formulas in both the
Relaxation Time Approximation (RTA) and the Chapman-Enskog approximation (CE). 
We investigate  and emphasize the differences between the isotropic and anisotropic 
cross sections and between the massless and massive particles. 
We show that in the range of temperature explored in a Heavy Ion collision and for pQCD-like cross section the 
RTA significantly underestimates the viscosity by about a factor of  2-3, 
while  a good agreement is found between the CE approximation and Gree-Kubo relation already at first order of 
approximation. 
The agreement with the CE approximation supplies an analytical formula that
allows to develop kinetic transport theory at fixed shear viscosity to entropy density ratio, $\eta/s$. This open the possibility
to explore dissipative non-equilibrium evolution of the distribution functions vs  T-dependent $\eta/s$ and particle momenta 
in the dynamics of the  Quark-Gluon Plasma created in ultra-relativistic heavy-ion collisions.

\end{abstract}

\section{Introduction}
The dynamics of the matter created in heavy-ion collisions (HIC) at both
the RHIC program at BNL and the LHC one at CERN
has shown that the azimuthal asymmetry in momentum 
space, namely the elliptic flow $v_{2}$, is the largest ever seen in HIC \cite{STAR_PHENIX}.
The comparison of the experimental measured $v_{2}$ with hydrodynamical calculations 
has suggested that in these HICs an almost perfect fluid with a very small shear 
viscosity to entropy density ratio $\eta/s$ has been created \cite{Romatschke:2007mq,Heinz}. Similar 
conclusions has been obtained also by kinetic transport theory \cite{Xu:2007jv,Xu:2008av,Ferini_PLB, Greco:2008fs} . At  
the level of comparison performed till now all these approaches have shown an agreement 
on the evaluation of the viscosity with $4\pi \eta/s \sim 1 -3$.
On the other hand the differential 
elliptic flow $v_2(p_T)$ has a very little change from the $\sqrt{s_{NN}}=39\,\rm GeV$
up to the last LHC data at $\sqrt{s_{NN}}=2.76 \, \rm TeV$.
It is an open question if this means an equal $\eta/s$ of the formed plasma between 
the RHIC and LHC energies or it is the result of different initial conditions and 
possible different non-equilibrium effects.

Both Hydrodynamical calculation and transport calculations have been widely used to study the dynamics of HIC
both showing that the elliptic flow depends sensitively on the ratio $\eta/s$.

Kinetic transport theory starts from a microscopic description of the  
dynamical evolution and hence needs the knowledge of cross section and 
mean fields, if Vlasov-like drifting terms are included \cite{Plumari_njl}.

Hydrodynamics instead is a macroscopic approach 
with dynamics ideally constrained by the conservation of energy-momentum tensor
and currents. Such oversimplification is possible only if dissipation is negligible,
while in realistic cases an expansion in terms of dissipative terms is needed.
This leads to relativistic viscous hydrodynamics usually developed according to the 
Israel-Stewart Theory \cite{Romatschke:2007mq,Heinz}.
Here viscosity represent an extrinsic parameter input to be calculated by the pertinent
quantum field theory or related effective lagrangian approximation.

In kinetic transport theory dissipative dynamics due to finite viscosity  is instead intrinsically 
included due to the presence of finite cross sections. However usually kinetic theory
is applied to the study of HIC starting from the microscopic details of the fields
and cross sections and it is not discussed directly in terms of viscosity of the system.
The search for the QGP properties, however, have shown that the shear viscosity and
in particular the viscosity to entropy density ratio $\eta/s$ is a key transport coefficient
that could be very close to the conjectured lower bound limit, $\eta/s= 1/4\pi$.
This has lead more recently to develop a transport approach at fixed $\eta/s$ 
\cite{Ferini_PLB,Molnar_cascade,Plumari_njl}
that allows to have a direct link
to the viscous hydrodynamic language. On the other hand kinetic theory is a tool than can allow to investigate
the non-equilibrium and dissipative effect in a wider range of validity respect to 
hydrodynamics for both $\eta/s$ and the momenta of the particles.
First attempts in this direction have been already developed and applied to 
the study of the QGP dynamics using the simple expression $\eta/s = 4T/(5\sigma_{tr})$, where
$\sigma_{tr}$ is the transport cross section.
However such an approach ask for a knowledge of the correct relation between the shear viscosity
$\eta$ and temperature, cross section, mass and density.

In literature there are several methods for the calculation of the shear viscosity,
the most employed ones are the Relaxation Time Approximation (RTA) \cite{reif}
and the Chapmann-Enskog (CE) method \cite{Groot}. The first is based on an ansatz for the collision integral in the 
Boltzmann equation and it does not allow to have control of the precision of the approximation.
The CE approach  is instead a variational approach that in principle allows to obtain 
solutions with an arbitrary accuracy which depends on the order of approximation used, \cite{Prakash:2012},
at variance with the RTA where it is not possible to have control over the degree 
of accuracy of the method.
Nonetheless the RTA approach has been even more often used due to its simplicity to evaluate
the viscosity for both hadronic and partonic matter \cite{Danielewicz:1984ww,Sasaki:2008fg,Sasaki:2008um,Thoma:1991em}.

In this work, we use the Green-Kubo relation \cite{GK} to extract the shear viscosity of a gluon gas from 
microscopic transport calculations within a parton cascade model with elastic two body collisions.
The Green-Kubo method in the contest of a relativistic plasma have been discussed also 
by other groups for the hadronic sector \cite{Demir_Bass,Muronga} 
and also in the QCD sector employing the pQCD matrix elements for $2\to 2$ collisions \cite{Fuini_3}
and more recently also for $2\to 2$ and $2\to3$ collisions \cite{Wesp_2011}.
An alternative method based directly on the stationary velocity gradients has been developed 
showing an excellent agreement with the Green-Kubo method \cite{Reining:2011xn}.

Our objective here is however quite different and it is essentially double folded. 
From one hand we want to compare the two main analytical approximation  schemes, CE and RTA, with the 
results obtained evaluating the Green-Kubo correlator solving it
numerically by the kinetic transport equation in a box with the test particle method.
From the other hand we will show that for all the case of interest the CE already at first order is 
a pretty good approximation to the exact viscosity. Hence providing an analytical
relation between $\eta \leftrightarrow T,\sigma(\theta),\rho, M$. Such analytical relation
supplies a way to construct a kinetic theory at fixed $\eta/s(T)$ with a much larger accuracy
respect to first tentatives \cite{Ferini_PLB,Molnar_cascade,Plumari_njl}, especially for the case
of non-isotropic cross section and massive quasi-particles both of interest for a realistic description
of both  quark-gluon plasma and hadronic matter.

%Such a comparison will be done for both isotropic and non-isotropic pQCD-like cross sections, as well as
%for massless and massive partons in a wide range of temperature of interest for the QGP physics.
In particular, we will show that the RTA approximation usually employed to evaluate the viscosity in NJL or quasi-particle models 
\cite{Sasaki:2008um,Khvorostukhin:2010aj,Khvorostukhin:2011mt,Gavin:1985ph},
$\eta/s=4\, T/ (5\,\sigma_{tr})$, can lead to large inaccuracy in the $\eta$ evaluation of more than a factor 2,
while the CE, already at first order, shows a satisfying  agreement with the Green-Kubo results.

The paper is organized as follows. In Section II, we discuss the method for calculating the Green-Kubo correlator
paying particular attention to the problem of convergency. In Section III, we give an overview of Chapmann-Enskog
and relaxation time approximation discussing the comparison with the Green-Kubo method for the cases
of isotropic and anisotropic cross section and for massive and massless particles. In Section IV, we apply our
method to the specific case of a gluon plasma with pQCD-like cross section that have a Debye screening mass
in the gluon propagator and we show that also in this physical case the relaxation time approximation can
significantly underestimate the $\eta/s$ of the plasma. Finally Section V contains summary and conclusions.

\section{Shear viscosity from Green-Kubo relation}

The transport coefficients like heat-conductivity, bulk and shear viscosity can be related to the correlation functions
of the corresponding flux or tensor in thermal equilibrium \cite{GK}. The underlying physical reason is that dissipation
of fluctuations have the same physical origin as the relaxation towards equilibrium, hence both dissipation and relaxation
time are determined by the same transport coefficients.Here we are interested to the shear viscosity $\eta$ for which 
the Green-Kubo formula assume the following expression \cite{zubarev}:

\begin{equation}
\label{green-kubo-completa}
\eta=\frac{1}{T} \int_{0}^{\infty}dt\int_{V}d^3x\,\langle \pi^{xy}(\textbf{x},t)\pi^{xy}(\textbf{0},t) \rangle
\end{equation}
where $T$ is the temperature, $\pi^{xy}$ is the $xy$ component of the shear component of the energy momentum tensor 
while $\langle \,...\,\rangle $ denotes the ensemble average.
In this work we determine numerically the correlation function $\langle \pi^{xy}(t)\pi^{xy}(0) \rangle$ solving the 
ultra-relativistic Boltzmann transport equation.

The correlations of shear components can be computed by employing transport simulations for a particle system
in a static box of volume V at equilibrium. We perform such simulation by mean of a relativistic transport
code already developed to perform studies of the dynamics of heavy-ion collisions at both
RHIC and LHC energies \cite{Ferini_PLB,Greco:2008fs,Plumari_BARI,Plumari:2010fg,Plumari_njl}.
The parton cascade developed solves a relativistic Boltzmann-Vlasov equation:

\begin{eqnarray}
\label{VlasovNJL}
p^{\mu}\, \partial_{\mu} f(x,p)+M(x)\partial_{\mu} M(x) \partial_{p}^{\mu} f(x,p)=\mathcal{C}(x,p)
\end{eqnarray}
where $f(x,p)$ is the distribution function for on-shell particles and $\mathcal{C}(x,p)$ is the Boltzmann-like collision integral that for a one component system can be written in a compact way as,

\begin{equation}
\label{coll}
\mathcal{C}(x,p)\!=\! 
\int\limits_2\!\!\! \int\limits_{1^\prime}\!\!\! \int\limits_{2^\prime}\!\!
 (f_{1^\prime} f_{2^\prime}  -f_1 f_2) \vert{\cal M}_{1^\prime 2^\prime \rightarrow 12} \vert^2 
 \delta^4 (p_1+p_2-p_1^\prime-p_2^\prime)
\end{equation}
where $\int_j= \int d^3p_j/(2\pi)^3\, 2E_j $, ${\cal M}$ denotes the transition matrix for the elastic processes and $f_j$ are the particle distribution functions, directly linked to the differential cross section $|{\cal M}|^2=16 \pi\,s\,(s-4M^2) d\sigma/dt$ with $s$ the Mandelstam invariant.
Indeed the calculations of the transport coefficients, based on the formal correspondence between the kinetic theory
and the hydrodynamics, are carried out as approximations of the collision integral in Eq.(\ref{coll}), 
see ref.s \cite{Prakash:2012,Moroz:2011,Kapusta_qp,Muronga_0,Denicol:2012es} .

Our aim here is to solve numerically the full collision integral to evaluate the viscosity through the Gree-Kubo
formula and compare it with the results of the RTA and CE approximation scheme.
The particle dynamics is simulated via Monte Carlo methods based on the stochastic interpretation of transition \cite{Xu:2004mz,Ferini_PLB},
according to which collision probability of two particle in a cell of volume $\Delta V_{cell}$ and within a time step $\Delta t$ is
\begin{equation}
P=v_{rel} \frac{\sigma_{tot}}{N_{test}} \, \frac{\Delta t}{\Delta V_{cell}}
\end{equation}
where $\sigma_{tot}$ is the total cross section and $v_{rel}=\sqrt{s(s-4 M^2)}/2\,E_1\,E_2$ denotes the relative velocity of the two incoming particle.
This method has been shown to be appropriate also for collisions between ultra-relativistic massless particles,
once one choose a sufficiently small time step $\Delta t$ respect to the time scale of the process to be studied, in our case
we have chosen $\Delta t= 0.01 \,\rm fm/c $.
We have performed calculation in a stationary box of volume $ V_{box}= 27 \,\rm fm^3$ with periodic boundary condition 
and a grid cell volume $\Delta V_{cell}=(0.1)^3 \, \rm fm^3$, checking that further increase of $V_{box}$ and reduction of
$\Delta V_{cell}$ does not change the results. The number of test particle $N_{test}$ allows to reduce spurious statistical
fluctuations, usually it is chosen to reproduce the behavior of known quantities such as the energy density $\epsilon$ and 
the pressure $P$. We have of course checked that these quantities are correctly reproduced, 
however in our case it is need a much more careful study of the convergency toward the exact solution
because we are interested to study temporal correlations and not simply global thermodynamical observable, see below for 
a more detailed analysis of this point.

The shear component of the energy momentum tensor is given by
\begin{equation}\label{pixy_generale}
\pi^{xy}(\textbf{x},t)=T^{xy}(\textbf{x},t)=\int \frac{d^3p}{(2\pi)^3} \frac{p^xp^y}{E}f(\textbf{x},\textbf{p};t)
\end{equation}
where we notice that at equilibrium the shear stress tensor is given by the energy-momentum tensor.
In our calculation the particles are distributed uniformly in the box. Therefore for an homogeneous system the volume averaged shear tensor
can be written as
\begin{equation}
\pi^{xy}(t)=\frac{1}{V}\sum_{i=1}^{N}\frac{p^x_i p^y_i}{E_i}
\end{equation}
the sum is over all the particles in the box.

%To determine numerically the correlation function, the last equation is written as follows
Numerically it is useful to write the correlation function $\langle \pi^{xy}(t)\pi^{xy}(0) \rangle$ as follows
\begin{eqnarray}
\langle \pi^{xy}(t)\pi^{xy}(0) \rangle = \left\langle \lim_{T_{max}\to\infty}\frac{1}{T_{max}}\int_{0}^{T_{max}}dt'\,\pi^{xy}(t+t')\pi^{xy}(t') \right\rangle =\\
\nonumber = \left\langle \frac{1}{N_{T_{max}}}\sum_{j=1}^{N_{T_{max}}}\,\pi^{xy}(i\Delta t+j\Delta t)\pi^{xy}(j\Delta t) \right\rangle
\end{eqnarray}
where $T_{max}$ is the maximum time to choose in our simulation, $N_{T_{max}}=T_{max}/\Delta t$ 
is the maximum number of time steps and $i\Delta t=t$, while $\langle\,\,...\,\rangle$ 
denotes the average over events generated numerically.

\begin{figure}
 \centering
 \includegraphics[scale=0.25, keepaspectratio=true]{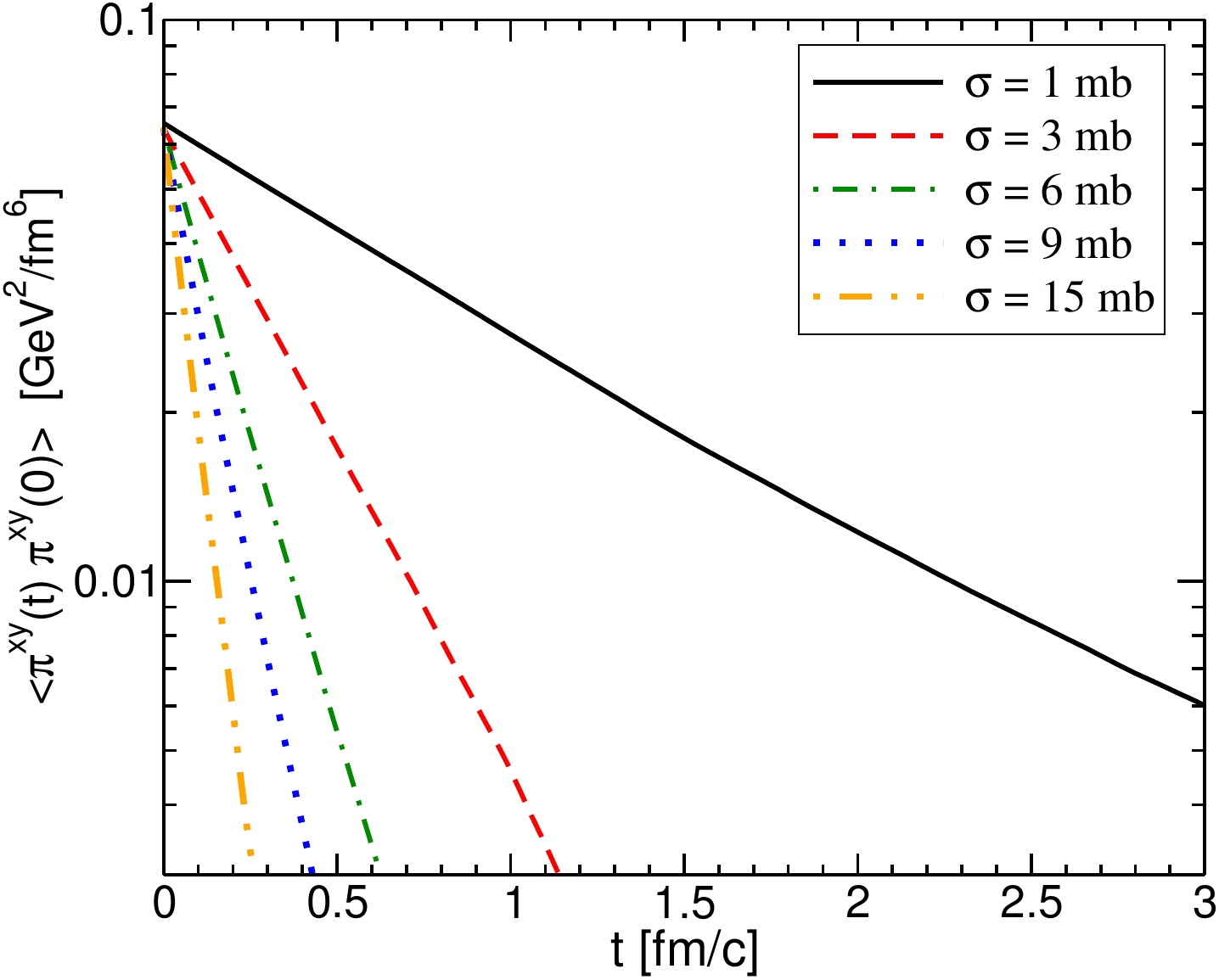}
 \caption{(Color online) Correlation function $\langle \pi^{xy}(t)\pi^{xy}(0) \rangle$ as a function of time for different value 
of the total cross section $\sigma_{tot}$, the temperature is fixed to $T=0.4 \, GeV$.
 These results are for massless particles and for isotropic cross section.}
 \label{fig:correlator}
\end{figure}

\begin{figure}
 \centering
 \includegraphics[scale=0.25, keepaspectratio=true]{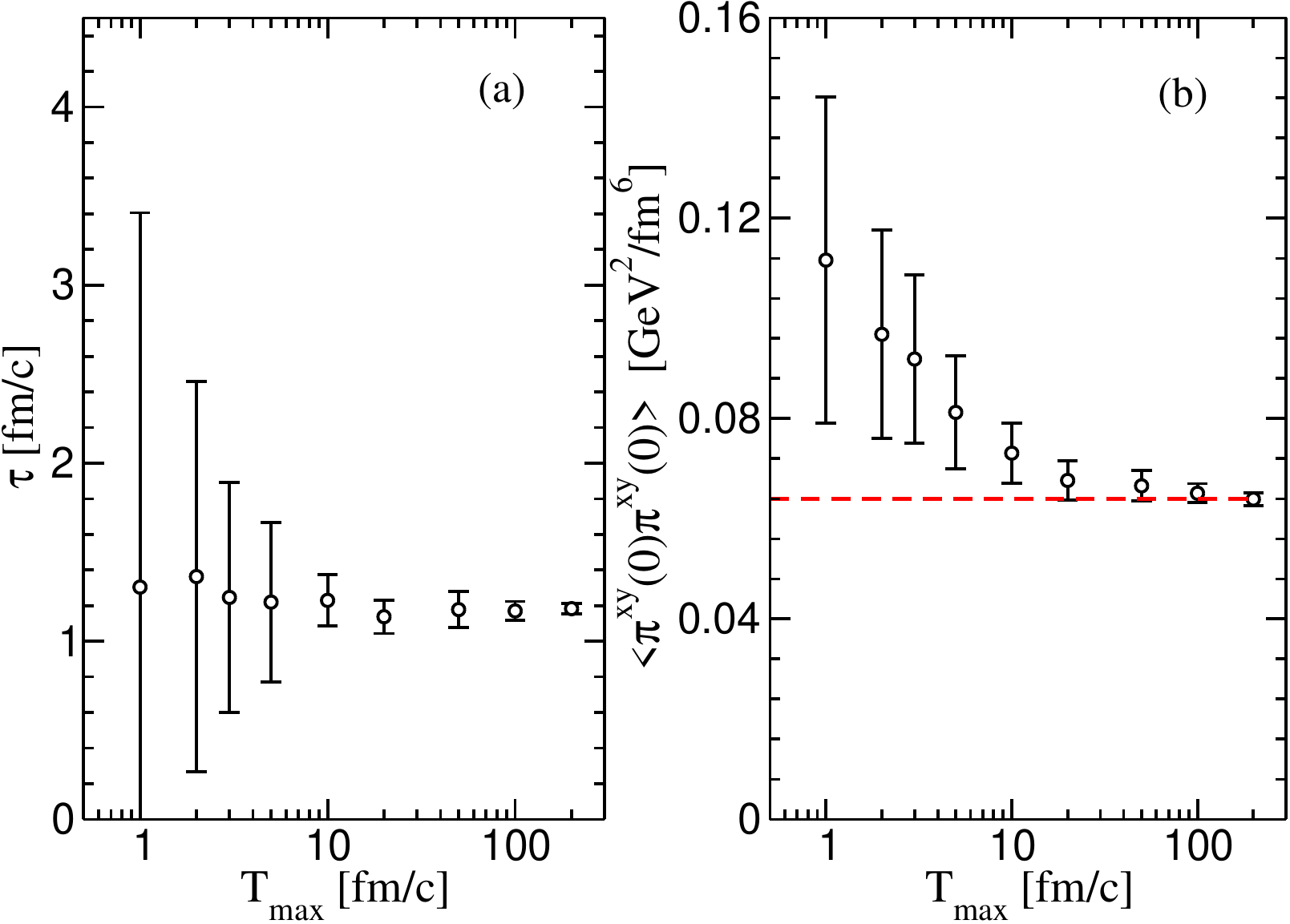}
 \caption{Left: relaxation time $\tau$ as a function of the maximum time $T_{max}$. Right: initial value of 
    the correlator $\langle \pi^{xy}(0)\pi^{xy}(0) \rangle$ as a function of $T_{max}$. The total cross section 
  is fixed to $\sigma_{tot}=0.1\,\rm fm^{2}$ and the temperature is $T=0.4 \, \rm GeV$. The dashed line
  is the analytical result.}
 \label{fig:corr_conver}
\end{figure}

In Fig. (\ref{fig:correlator}) it is shown the correlation function $\langle \pi^{xy}(t)\pi^{xy}(0) \rangle$ 
as a function of time for different values of the total cross section $\sigma_{tot}$. In these calculations the particles are 
distributed uniformly in the box in coordinate space, while in the momentum space we have considered a thermal distribution
($f \sim e^{-E/T}$) (at a temperature of $T=0.4 \,\rm GeV$). In the particular case shown in Fig.(\ref{fig:correlator}) the particles 
are massless and they interact with an isotropic energy independent differential cross section.
As we can see the function $\langle \pi^{xy}(t)\pi^{xy}(0) \rangle$ is an exponential decreasing function. 
We notice that the exponential decay with time is not an assumption, but it comes directly
from the calculation from the dynamical evolution in the simulation.
We use 
this fact to fit the correlation function with the following expression, as done in several other works \cite{Wesp_2011,Fuini_3,Demir_Bass,Muronga},
\begin{eqnarray}
\langle \pi^{xy}(t)\pi^{xy}(0) \rangle = \langle \pi^{xy}(0)\pi^{xy}(0) \rangle e^{-t/\tau}
\label{corr_fit}
\end{eqnarray}
$\tau$ is the so called relaxation time. As we can see in Fig.(\ref{fig:correlator}), the initial point of the 
correlation function is independent on the strengh of interaction but it depends only on the temperature and 
volume of the system.
This result found numerically is indeed expected because it is possible to show that the initial value of the correlation 
function can be expressed by the following relation
\begin{eqnarray}
\langle \pi^{xy}(0)\pi^{xy}(0) \rangle= \frac{4}{15} \frac{e T}{V}
\label{pi00}
\end{eqnarray}
from which we can see that the correlator at $t=0$ depends only on thermodynamical variables and not on the cross section.
We have found an agreement between the analytical value in Eq.(\ref{pi00}) and the numerical result at the level of $1\%$,
see Fig. (\ref{fig:corr_conver}).

The slope of the correlation function (i.e. $\tau$) depends on the scattering cross section and we will show that as expected
 from kinetic theory $\tau = c/(\sigma \rho)$ with the coefficient $c$ that depends on the mass of the particles (or better on the $m/T$ ratio,
 see next Section)
and on the angular dependence of the scattering cross section. 
Substituting Eq. (\ref{corr_fit}) into Eq. (\ref{green-kubo-completa}) the formula for the shear viscosity 
becomes:
\begin{equation}
\eta=\frac{V}{T}\langle \pi^{xy}(0)\pi^{xy}(0) \rangle \tau
\end{equation}
This is the formula that we will use in our calculation to extract the shear viscosity.
The relaxation time $\tau$ is calculated performing a fit on the temporal range where the correlation 
function assume the exponential form, because at $t >> \tau$ the correlation becomes too weak and the fluctuations
starts to dominate. A key point is the evaluation of the error on the value of the viscosity as coming from
the error on the initial value 
of the correlator  and the error on the relaxation time $\tau$ extracted from the fit of the correlation function, hence
possible deviation from the exponential law are evaluated through the error bars themselves.

In Fig. (\ref{fig:corr_conver}) it is shown an example of study of the convergence of the relaxation time $\tau$ and the initial 
value of the correlator $\langle \pi^{xy}(0)\pi^{xy}(0) \rangle$ as a function of the maximum time of the 
simulation $T_{max}$ for the case of isotropic and constant total cross section of $\sigma_{tot}=0.1\, \rm fm^{2}$ and for a 
temperature of $T=0.4 \, \rm GeV$. 
We have performed such analysis using a large number of test particles $N_{test}=1000$, the convergency
with $N_{test}$ will be described soon after.
As shown in Fig. \ref{fig:corr_conver} (right) $\langle \pi^{xy}(0)\pi^{xy}(0) \rangle$ converges to the analytical 
value given by Eq.(\ref{pi00}) (dashed line) with
good accuracy only at $T_{max} \approx \rm100 \,\rm fm/c$ , which is a very large time scale compared respect to the relaxation time 
for this case, $\tau \sim 1 \,\rm fm/c$.
The $\tau$ itself is instead always quite close to the exact value, but for small $T_{max}$, however, the 
evaluated error bars show that there is a large uncertainty on the exponential fit that again is reduced  increasing
the maximum time over which the temporal correlations are followed.
The large error bars are essentially indicating that for small $T_{max}$ it is not possible to have a defined
exponential decay behavior according to Eq.(\ref{corr_fit}).

\begin{figure}
 \centering
 \includegraphics[scale=0.25, keepaspectratio=true]{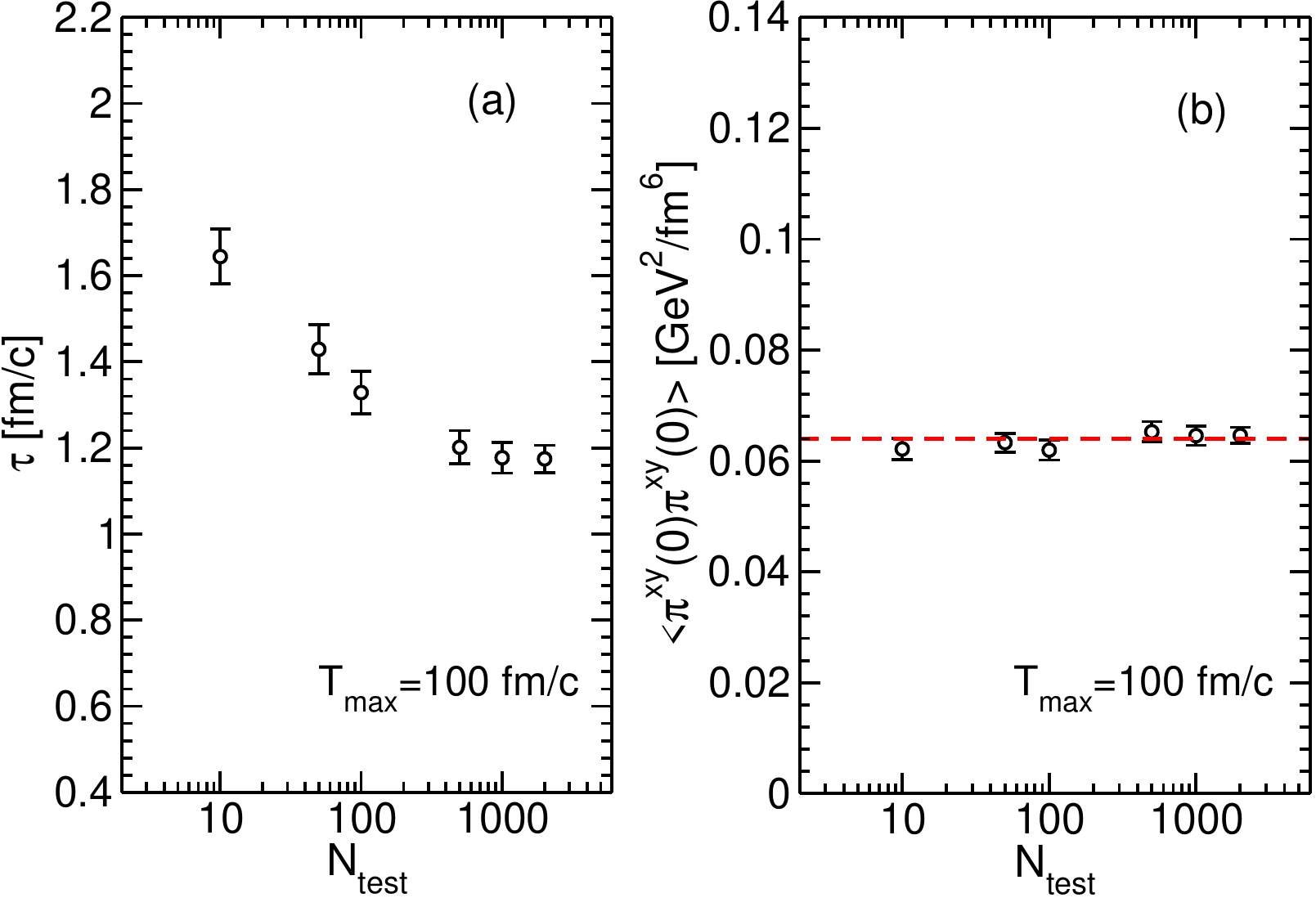}
 \caption{Left: relaxation time $\tau$ as a function of the maximum time $N_{test}$. Right: initial value of 
    the correlator $\langle \pi^{xy}(0)\pi^{xy}(0) \rangle$ as a function of $N_{test}$. The total cross section 
  is fixed to $\sigma_{tot}=0.1\, fm^{2}$ and the temperature is $T=0.4 \,\rm GeV$. The dashed line
  is the analytical result.}
 \label{fig:corr_conver_ntest}
\end{figure}

We have checked also the convergency vs the number of test particles, in Fig.(\ref{fig:corr_conver_ntest})
we see that for $\tau$ it is important to have a large number of test particle
$N_{test} \ge 500 $ to reduce the uncertainty on $\tau$ at the level of $2 \%$.
On the right panel of Fig.\ref{fig:corr_conver_ntest} the $\langle \pi^{xy}(0)\pi^{xy}(0) \rangle$
is shown to be nearly indipendent on $N_{test}$, as one can expect considering it is the initial
value of the correlation hence less affected by the accuracy of the system evolution.
At variance with the dependence of $\tau$ on $T_{max}$ the $N_{test}$ is of crucial importance.
This can be understood because a small number of $N_{test}$ does not allow to properly map
the phase space and consequently the dynamical evolution of the system.

A similar study have been performed for all the numerical calculations shown in the paper, we have used
$T_{max}=100\, \tau$, $N_{test}=1000$ and $\Delta t= 0.01\, \rm fm/c$.

\section{Chapman-Enskog and Relaxation Time Approximation vs Green-Kubo}

In literature there are several methods for the calculation of the 
shear viscosity, one of these is the Chapman-Enskog (CE) approach \cite{reif,Groot,zubarev} and 
another is the relaxation time approximation (RTA) \cite{Gavin:1985ph,Kapusta_qp}. The difference between 
these two methods resides in the different way in which the collision 
integral is approximated.
In the RTA is not possible to have control 
over the degree of accuracy of the method and furthermore it is not possible 
to go to higher order.
In the CE approximation instead  is possible to obtain solutions 
with an arbitrary accuracy which depends on the order of approximation used. We show briefly the two different approximation methods and 
formulas for the viscosity that we will use to compare the numerical results 
obtained using the Green-Kubo relation.

The starting hypothesis of the RTA is that the collision integral can be approximated by
\begin{eqnarray}
C[f]=-\frac{f-f^{eq}}{\tau}
\end{eqnarray}
where $\tau$ is the so called relaxation time. In this approximation we are assuming that the 
effect of the collision between the particles is to keep locally the system close to the equilibrium.
Collisions are assumed to restore the local equilibrium always with a relaxation time $\tau$
that is of the order of the time between two collisions. Hence the
the distribution function $f$ tends to the equilibrium value $f^{eq}$ exponentially with 
a time $\tau$. In this approach it has been demonstrated that the shear viscosity assumes 
the following expression \cite{Gavin:1985ph,Kapusta_qp}:

\begin{eqnarray}
 \eta &=& \frac{1}{15T}\,\int_0^{\infty}\,\frac{d^3p_a}{(2\pi)^3}\,
\frac{|p_a|^4}{E_a^2}\,\frac{1}{w_a(E_a)}\,f^{eq}_a\,\,
\label{eta_relax}
\end{eqnarray}
where $w_a(E_a)$ is the so called collision frequency and $f_a^{eq}$ is the
equilibrium distribution function of particles $a$ with momenta $p_a$
and energy $E_a$.
The relaxation time $\tau_{a}=\omega(E_{a})^{-1}$ is given by
\begin{equation}
\tau_{a}^{-1}=w_a(E_a) = \sum_{bcd} \frac{1}{2} \int \frac{d^3p_b}{(2\pi)^3} \,
\frac{d^3p_c}{(2\pi)^3} \,
\frac{d^3p_d}{(2\pi)^3} \, W(a,b|c,d)\, f^{eq}_b\,\,,
\label{tau_eq}
\end{equation}
where the quantity $W(a,b|c,d)$ is defined as
\begin{equation}
 W(a,b|c,d) = \frac{(2\pi)^4 \delta^4\,(p_a+p_b-p_c-p_d)}
{2E_a2E_b2E_c2E_d} \,|\mathcal{M}|^2\,\,.
\end{equation}
$|\mathcal{M}|^2$ is the squared transition amplitude for the 2-body
reaction $a+b \rightarrow c+d$. The collision frequency Eq.(\ref{tau_eq}) 
can easily be expressed in terms of the total cross section $\sigma_{tot}$:

\begin{eqnarray}
  w_a(E_a)  &=&  \int \frac{d^3p_b}{(2\pi)^3}\, \,\frac{\sqrt{s(s-4m^2)}}{2E_a\,2E_b} \,f^{eq}_b\,\sigma_{tot}\,\,,
  \label{tau_eq_2}
\end{eqnarray}

that for a constant (energy independent) total cross section becomes energy independent and 
coincides with the standard mean relaxation time

\begin{eqnarray}
 \tau_a^{-1}=w_a(E_a)=\rho \, \sigma_{tot} \, \langle v_{rel} \rangle
  \label{tau_eq_4}
\end{eqnarray}
where $\langle v_{rel} \rangle$ is the thermal average of the relative velocity and for massless particles 
$\langle v_{rel}\rangle=1$. 
We notice that in the RTA the interaction appears 
in the collision frequency only through the total cross section. However on general physical argument the 
viscosity is expected to depend also on the momentum transfer that on average the collisions are able
to produce. Therefore one would expect that different angular dependent cross sections generate a
different viscosity even if they have an equal total cross section that is associated only to the probability
to collide and not to momentum transfer occuring in the collisions. 
To occur for the $q^2-$transfer efficiency  of the cross section, in transport theory one defines the transport cross section,
$\sigma_{tr}=\int d(cos \theta) \sigma(\Theta) (1-cos^2\Theta)$, i.e.
the differential cross section weighted by the momentum transfer that is proportional to ($1-cos^2\Theta$).
In the literature sometimes to take into account this fact the relaxation time is approximated 
by $\tau^{-1}_{tr}=\langle \rho \, \sigma_{tr} \, v_{rel} \rangle$, i.e. substituting the total with the transport
cross section. This is not really coming from the RTA as in Eq.s (\ref{eta_relax}) and (\ref{tau_eq_2}), however
we will also refer to it as modified RTA in the following, but we will see that also such an extension of the RTA
can reasonably approximate the correct viscosity only for the case of isotropic cross section, in which case 
$\sigma_{tr}=(2/3) \sigma_{tot}$.

To describe the CE approach we use the formalism recently developed in Ref. \cite{Prakash:2012}, where
it is discussed how the shear viscosity at first order $[\eta_s]_I$ can be written for the
most general case of relativistic particles at finite mass colliding with a non-isotropic and energy-dependent cross section as 
\begin{eqnarray}
 [\eta_s]_I &=&\frac{T}{10}\,\frac{\gamma_0^2}{c_{00}}
\label{shear_I}
\end{eqnarray}

where $ \gamma_0 = -10 \hat{h} $, with $z ={m}/{T}$ and $\hat{h} = K_3(z)/K_2(z)$ is the enthalpy and 
\begin{equation}
 c_{00} = 16\left( w_2^{(2)} - \frac{1}{z}\,w_1^{(2)} + \frac{1}{3z^2}w_0^{(2)} \right) \label{c00massive}
\end{equation}

The $w_i^{(s)}$ are the so-called relativistic omega integrals which are defined as
\begin{eqnarray}
w_{i}^{(s)} &=& \frac{2\pi z^3}{K_2(z)^2}\int_{1}^{\infty} dy\, y^i 
(y^2-1)^3 \, K_j(2zy)\nonumber\\
& & \times\,\int_{0}^{\pi} d\Theta \, \sin\Theta \, \sigma(s,\Theta) \, (1-\cos^s\Theta)~.
\label{omega_integral}
\end{eqnarray}
%\begin{eqnarray}
%w_{i}^{(s)} &=& \frac{2\pi z^3}{K_2(z)^2}\int_{0}^{\infty} d\psi
%\sinh^7\psi \, \cosh^i\psi \, K_j(2z\cosh\psi)\nonumber\\
%& & \times\,\int_{0}^{\pi} d\Theta \, \sin\Theta \, \sigma(\psi,\Theta) \, (1-\cos^s\Theta)~.
%\label{omega_integral}
%\end{eqnarray}
where $\sigma(s,\Theta)$ is the differential cross section and 
$j = \frac{5}{2}+\frac{1}{2}\left( -1\right)^i$ while $y = \sqrt{s}/2M$.
Note that the angular integration in the $\omega_{i}^{(s)}$ for $s=2$ is proportional to the 
transport cross section $\sigma_{tr}=\int \, d\Omega \, \sigma(s,\Theta) \, \sin^{2}\Theta$.

In the following we will consider the case of isotropic $\sigma$ for both massless and massive particles
and the more realistic case of non-isotropic cross section for massless  particles.

\subsection{Isotropic cross section: massless case}

In this section we compare our results for $\eta$ using the Green-Kubo relation and the analytical result 
obtained in the relaxation time approximation and the Chapman-Enskog approach, for the case of massless gluons \footnote{
In our result in this Section there is nothing specific of a gluon system and the results discussed are however very general.} 
colliding with an isotropic cross section $\sigma(s,\Theta)=\sigma_{0}=const$.
The CE first order approximation for the shear viscosity, Eq.(\ref{shear_I}), for the case considered is simply:
\begin{eqnarray}
[\eta]^I_{CE}= 0.8 \frac{T}{\sigma_{tr}}=1.2 \frac{T}{\sigma_{tot}}
\end{eqnarray}
where for the last equation we have used $\sigma_{tr}=2/3 \sigma_{tot}$ valid only for isotropic cross section. 
We notice that in  CE already at first order of approximation appears the differential 
cross section weighted with ($1-cos^2\Theta$), i.e. the transport cross section. 
This comes from the $\omega_{i}^{(2)}$ integrals that independently on the $i-$value are proportional to $\sigma_{tr}(s)$,
hence one can re-write $c_{00}= f(z) \sigma_{tr}$, with $f(z)$ that can be written in terms of a combination
of Bessel functions with coefficients coming from the energy integration present in Eq.(\ref{omega_integral}).
So one finds in general $\eta \sim 1/\sigma_{tr}$ as expected.
In the next section we derive a more explicit formula for the case of a pQCD cross section with HTL-like dressed propagator.

For this simple case of isotropic cross section and massless particle in the literature there exist also higher order
calculation up to the most recent work in Ref. \cite{Prakash:2012} where the calculation was extended
up to the $16^{th}$ order. They found that higher order approximation converge to the value
\begin{eqnarray}
[\eta]^{\tiny 16^{th}}_{{\tiny CE}} = 0.845 \frac{T}{\sigma_{tr}}= 1.267 \frac{T}{\sigma_{tot}}
\label{Chapman_16}
\end{eqnarray}
where however approximations beyond the third order give corrections much smaller than $1 \%$, while the difference
between the $1^{st}$ and the $16^{th}$ order is about $6 \%$.
 
In the RTA the shear viscosity for the massless case can be calculated using 
Eq.(\ref{eta_relax}) and Eq.(\ref{tau_eq_4}) easily obtaining:
\begin{eqnarray}
\eta_{RTA} = 0.8 \frac{T}{\sigma_{tot}}
\end{eqnarray}
Hence the difference between the Chapman-Enskog (CE) at first order and the relaxation time approximation (RTA)
can be seen as the substitution of $\sigma_{tot}$ with $\sigma_{tr}$.
This result predict a viscosity about $40 \%$ smaller respect to the CE approximation.
As mentioned in the previous section the RTA does not take into account the transport cross section, but quite often
the RTA is used choosing a $\tau^{-1}=\langle \rho \, \sigma_{tr} \, v_{rel} \rangle$, where again for an energy and 
angular independent cross section one obtains the same results in first order CE:
\begin{eqnarray}
\eta^*_{RTA} = 0.8 \frac{T}{\sigma_{tr}}= 1.2 \frac{T}{\sigma_{tot}}.
\end{eqnarray}
where we put a $"^*"$ to indicate that it is not exactly the RTA but a sort of modified one.
We will show that this is correct only if the 
cross section does not depend neither on the energy nor on the scattering angle. In other terms it cannot be applied to 
the case of a QCD matter and more generally to hadronic and partonic effective lagrangian approach
that always predict quite non-isotropic scattering cross section.

\begin{figure}
 \centering
 \includegraphics[scale=0.25, keepaspectratio=true]{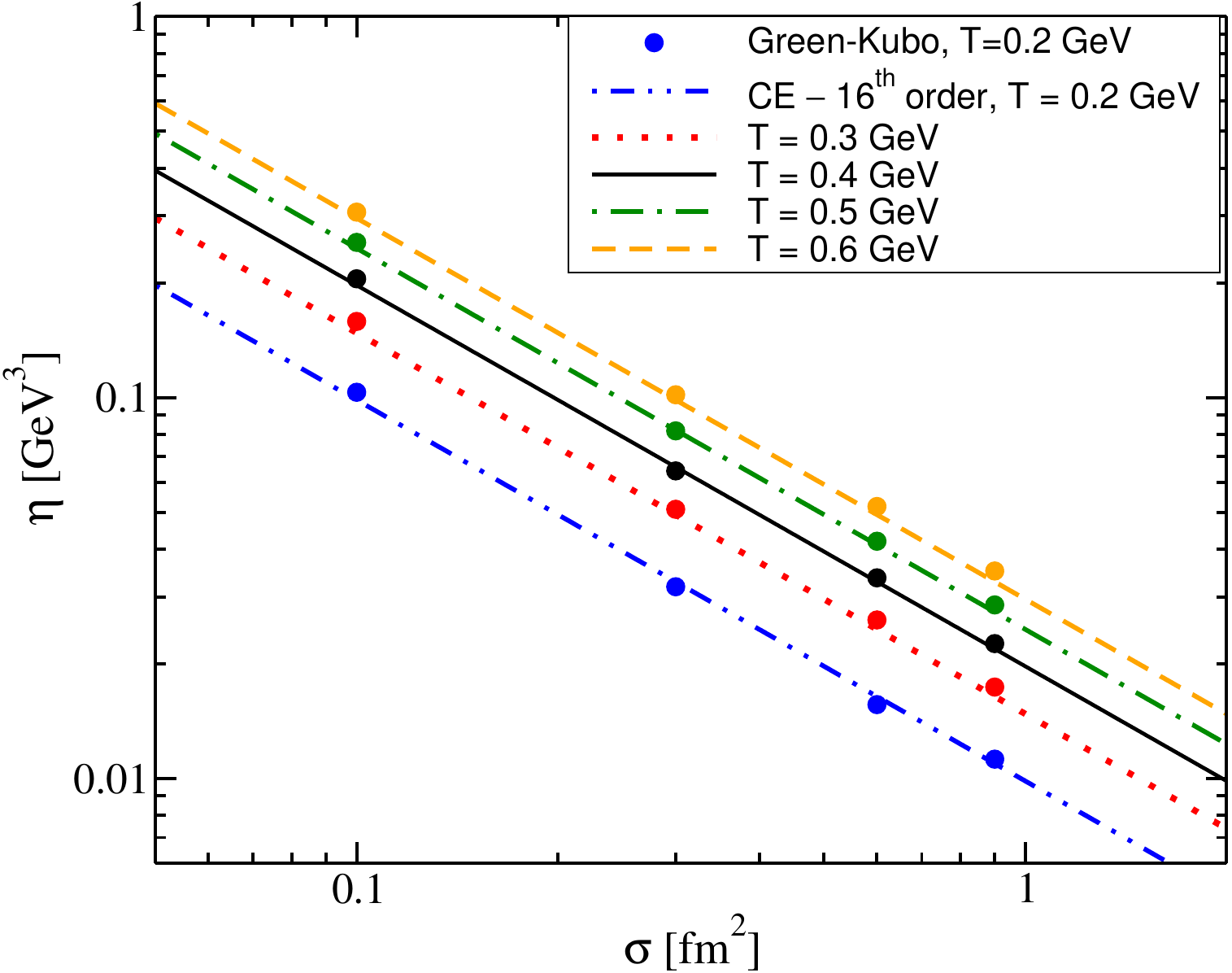}
 \caption{(Color online) Shear viscosity $\eta$ for a massless system as a function of the total isotropic cross section $\sigma$ and for different temperatures. 
The circles are the results from the Green-Kubo method, while the lines are the 
results obtained using the Chapman-Enskog approximation in Eq.(\ref{Chapman_16}).}
 \label{fig:eta_sigma}
\end{figure}

Our results obtained using the Green-Kubo formula are shown in Fig.(\ref{fig:eta_sigma}) by full circles
 and compared to the prediction of CE at $16^{th}$ order, Eq.(\ref{Chapman_16}) shown by different lines
 for each temperature from 0.2 to 0.6 GeV . 
The error bars for the Green-Kubo calculation are small and within the symbols.
 As we can see we have a very good agreement with the analytical 
results in all the examined range of cross sections and temperatures. Discrepancies are observed
only at the level of about $2\%$, which is comparable to the uncertainty 
in the numerical evaluation of the Green-Kubo correlator.

In our simulations the gluons are considered as massless particles and due to the fact that they are distributed
according to the Boltzmann distribution the entropy density is given by $s = 4 \rho = 4 d_g T^3 / \pi^2$ with $d_g=16$. 
We have checked that the transport code correctly reproduce the analytical formula by calculating
$\langle \Pi^{00}+\Pi^{xx} \rangle/T$. 
Therefore for a temperature independent cross section we have
\begin{eqnarray}
\eta/s = 0.195 \frac{1}{ \sigma_{tot} \, T^2}
\label{etas_eq}
\end{eqnarray}

\begin{figure}
 \centering
 \includegraphics[scale=0.25, keepaspectratio=true]{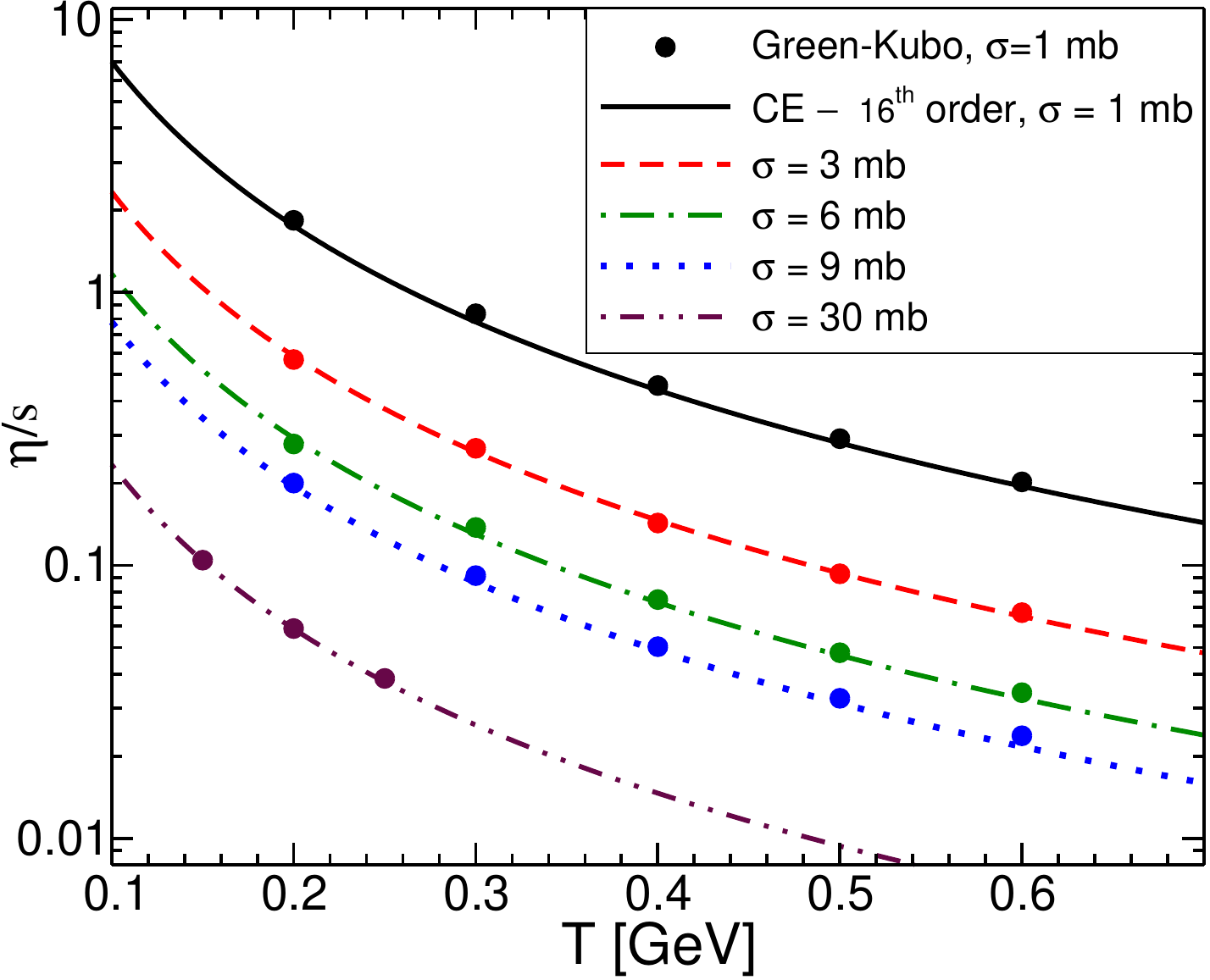}
 \caption{(Color online) $\eta/s$ as a function of the temperature and for different cross section $\sigma_{tot}= 1, 3, 6, 9, 30 \, mb$. 
The circles are the results of the Green-Kubo method while the lines are the result obtained using Eq.(\ref{etas_eq})
from CE approximation.}
 \label{fig:etas_T}		
\end{figure}

In Fig.(\ref{fig:etas_T}) the shear viscosity to entropy density ratio $\eta/s$ is shown as a function of the 
temperature and for different total cross section $\sigma$. The circles are the results of our simulations 
while the lines are the analytical results obtained using Eq.(\ref{etas_eq}).
From the value of $\eta/s$ we can see that we have explored the scale of $\sigma, \rho,T$ of interest for the study of the QGP.
The viscosity extracted 
with the Green-Kubo relation is in good agreement with the CE approximation in a very wide range of density and 
cross section. We notice that the results shown correspond to a gas condition with a mean free path $0.005\leq \lambda\leq 5.95\,\rm fm$ 
corresponding respectively to $\sigma= 30 \, \rm mb$ and $\rho= 45.6\,\rm fm^{-3}$ ($T=0.6\,\rm GeV$) and $\sigma= 1 \, \rm mb$ and $\rho=1.68\, \rm fm^{-3}$
 ($T=0.2\, \rm GeV$).
Sometimes in the literature it is stated that the Boltzmann transport equation is applicable only in a regime of diluteness.
However the meaning of such a statements is often misunderstood. It does not mean that one cannot study by mean of Boltzmann
transport equation the system dynamics if the mean free path is small respect to the average particle distance. In fact our work shows how one can fix the
cross section according to the CE approximation and the Green-Kubo formula, demonstrating that in such a way one really
has a fluid with the wanted viscosity even if this corresponds to very small mean free path respect to particle
distance. Of course what one cannot
deduce from this is if microscopically in the real system the viscosity comes only from two-body collisions.

\subsection{Isotropic cross section: massive case}

The calculations discussed till now consider only massless particles: an interesting 
extension is that  with finite mass.
The importance of this kind of calculation for a partonic system is that it can be employed to evaluate the 
viscosity in quasi-particle models \cite{Plumari_qpmodel,Ruggieri_qpmodel}.
We remind that  the quasi-particle approach is important because it allows to describe the equation of state obtained on the 
lattice of a plasma of quarks and gluons  \cite{Plumari_njl,Scardina_2012}. This is achieved incorporating the non-perturbative effects of 
QCD in a temperature dependent mass term for both gluons and quarks. 
Furthermore finite mass is relevant also if one wants to study the viscosity of an hadronic fluid or employ transport theory to study the dynamics
of hadronic matter the mass cannot be neglected.

\begin{figure}
 \centering
 \includegraphics[scale=0.25, keepaspectratio=true]{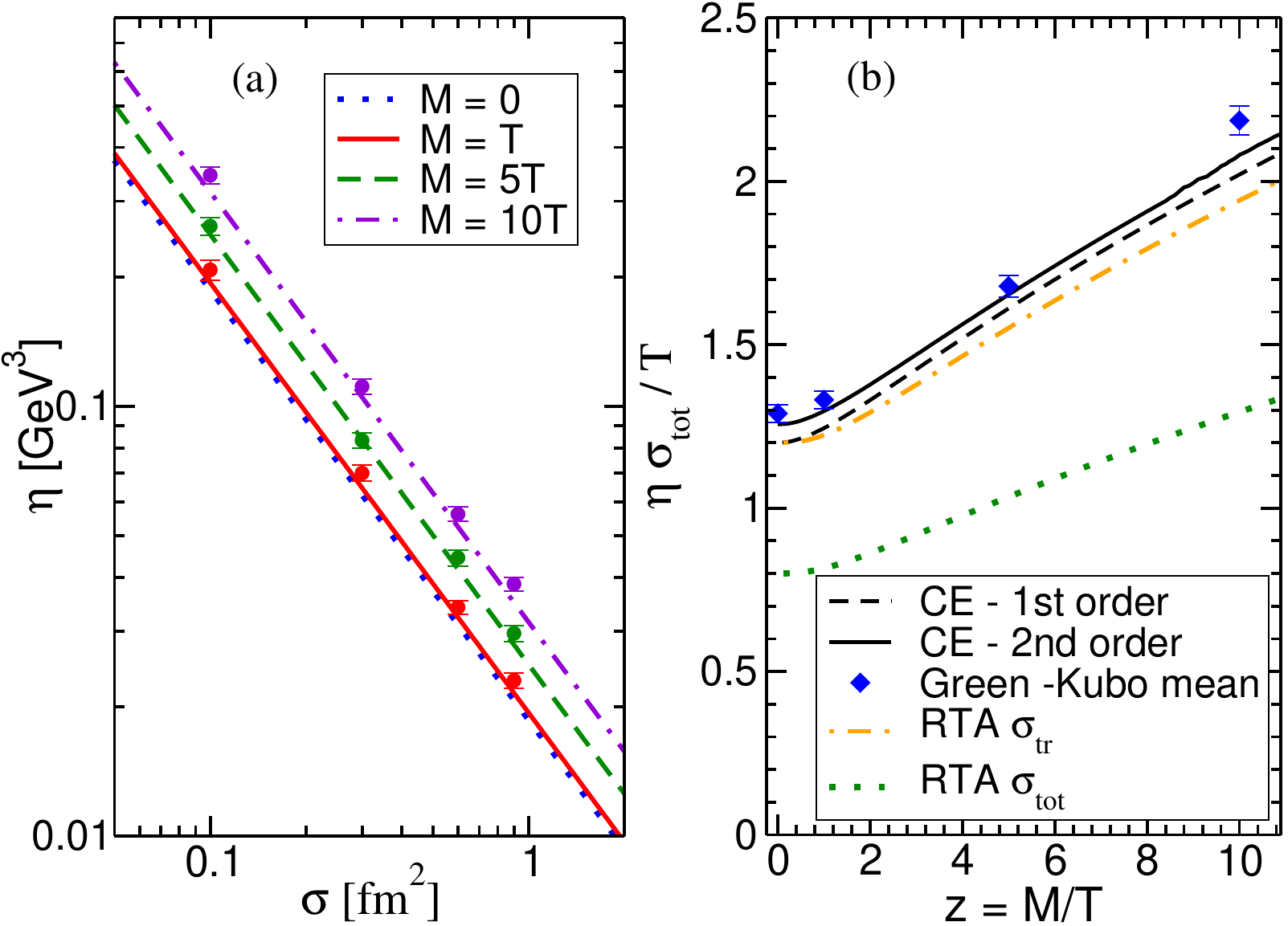}
 \caption{ (Color online) Left: Shear viscosity $\eta$ as a function of the Debye mass $m_{D}$ for three different values of the temperature $T=0.3, 0.4, 0.5 \, GeV$ blue, green and red respectively. The open simbols are the results obtained using the Green-Kubo relation.
The solid, dashed and dot dashed lines refer to the RTA approach with $\tau^{-1}=\langle \rho \, \sigma_{tr} \, v_{rel} \rangle$ respectively for $T=0.3, 0.4, 0.5 \, GeV$. The dotted line is the isotropic limit when $m_D \to \infty$. 
Right: The same as the left panel but the solid, dashed and dot dashed lines refer to the Chapman-Enskog approximation at first order.
}
 \label{fig:eta_M_T}		
\end{figure}

In the left panel of Fig.(\ref{fig:eta_M_T}) the results for the shear viscosity $\eta$ for the massive case, obtained using 
the Green-Kubo relation,  are shown and compared to the analytical result obtained using 
the CE approximation.
For a constant differential cross section $\sigma(s,\Theta)=\sigma_{0}$
the shear viscosity in the CE approach 
can be written at first order approximation, Eq.(\ref{shear_I}), in terms of the total cross section $\sigma_{tot}$ as
\begin{eqnarray}
\eta = f(z) \frac{T}{\sigma_{tot}}
\label{eta_M_eq}
\end{eqnarray}
where the function $f(z)$ \cite{Groot,Moroz:2011}
\footnote{In the De Groot  {\it et al.} book it is present a typos
in the denominator where 15 $z^2$ is written as 5$z^2$.}

\begin{eqnarray}
f(z) = \frac{15}{16}\frac{z^4 K_3^2(z)}
 {(15z^2+2)K_2(2z)+(3z^3+49z)K_3(2z)}
\end{eqnarray}

where $f(z) \to 1.2$ in the ultra-relativistic limit $z\to 0$. 
In the right panel $f(z)$ of Fig.(\ref{fig:eta_M_T})  is shown as a function of $z=m/T$ for the first order
(dashed line) and second order (solid line)  of approximation. The diamonds represent the mean value over the different
cross sections of the results obtained using the 
Green-Kubo relation for the viscosity $\eta$ for each value of mass explored. As we can see, we have a fairly good agreement 
with the analytical result already at first order. However as already seen for $M=0$ there is a discrepancy of about $6\%$
at first order corresponding to the difference between $\eta \, \sigma_{tot}/T=1.2$ and $1.267$. Here we have evaluated
numerically also the CE expansion at second order that according to \cite{Prakash:2012}: 

\begin{eqnarray}
\left[ \eta_s\right]_{II} &=& \frac{T}{10}\,
\frac{\gamma_0^2\,c_{11} - 2\,\gamma_0\,\gamma_1\,c_{01} + \gamma_1^2\,c_{00}}
{c_{00}c_{11}-c_{01}^2} \label{shear_2}
\end{eqnarray}
where
\begin{eqnarray}
 \gamma_1 &=& - \left[ \hat{h}(10z -25) -10z \right] \\
c_{01} &=& 8 \left[2z\left(w_2^{(2)} - w_3^{(2)} \right) + \left( -2w_1^{(2)} + 3w_2^{(2)}\right)\right. \nonumber \\
&& \left. + z^{-1}\left(\frac{2}{3}w_0^{(2)} - 9w_1^{(2)} \right) -\frac{11}{3z^2}w_0^{(2)} \right]
\end{eqnarray}

\begin{eqnarray}
c_{11} &=& 4\left[4z^2 \left(w_2^{(2)}-2w_3^{(2)} + w_4^{(2)}\right) \right. \nonumber \\
&& \left. + 2z \left(
-2w_1^{(2)}+ 6w_2^{(2)} - 9w_3^{(2)} \right) \right. \nonumber \\
&& \left. + \left(\frac{4}{3}w_0^{(2)} -36w_1^{(2)} +41w_2^{(2)} \right) \right. \nonumber \\
&& \left. + z^{-1}
\left(-\frac{44}{3}w_0^{(2)} - 35w_1^{(2)} \right) + \frac{175}{3z^2}w_0^{(2)}  \right]
\label{shearcoeffs}\nonumber \\
\end{eqnarray}

The behavior of $\left[ \eta_s\right]_{II} \sigma_{tot}/T$ is shown in Fig. (\ref{fig:eta_M_T}) (right) 
by solid line. We see that the second order CE improves the agreement with the Green-Kubo
results reducing the discrepancies to less than $3\%$. It seems that with increasing mass the 
convergency is slower and higher order are more relevant.

Our result shows that for massive particle the coefficient between $\eta$ and $T/\sigma_{tot}$
increases by about $40\%$ for value of $M/T$ typical of gluon mass in quasi-particle models
\cite{Plumari_qpmodel,LH1998,PC05,PKPS96} or protons around and below the cross-over 
temperature.
The comparison with the RTA shows also in this case that as we move from the simplest case
of massless particle the viscosity is underestimated. In the case of the pure RTA (dotted line) the
correct value, about a $40\% $ larger, become about a $70\%$ larger 
for a mass $M=10 T$, but also the modified RTA that is equal to CE at first order for $M=0$ 
under predict the $\eta$ when a finite mass is considered. The discrepancy increase
with the mass and it is about $15\%$ for $M=10 T$
even if the cross section is isotropic.

\subsection{Anisotropic cross section}

In this section our aim is to study the more realistic case of angular dependent cross section,
being in the context of a gluon plasma we choose typical elastic pQCD inspired cross section 
with the infrared singularity regularized by  Debye thermal mass $m_D$:
\begin{equation}
 \frac{d\sigma}{dt} = \frac{9\pi \alpha_s^2}{2}\frac{1}{\left(t-m_D^2\right) ^2}\left(1+\frac{m^2_D}{s}\right) 
\label{sigma_md}
\end{equation}
where $s,t$ are the Mandelstam variables. 
Such kind of cross sections are those typically used in parton cascade approaches 
\cite{Zhang:1999rs,moln02,Ferini_PLB,Greco:2008fs,Plumari:2010fg,Xu:2004mz,Xu:2008av}. 
The total cross section corresponding to Eq. (\ref{sigma_md}) is $\sigma_{tot}=9\pi \alpha_s^2/(2 m_{D}^2)$ 
which is energy and temperature independent. 
Here our objective is not to estimate the value of $\eta$ (see next Section) but
only to explore the impact of non-isotropic cross section on the comparison among RTA, CE and Green-Kubo
methods. In Eq.(\ref{sigma_md}) the Debye mass $m_D$ is a parameter that regulates the anisotropy of the 
scattering cross section. We will vary it to regulate the anisotropy, but fixing the total cross section
constant by keeping constant the $ \alpha_s^2/ m_{D}^2$ ratio.
We note that for a plasma at temperature $T$ the average possible momentum transfer $q^2 \approx (3T)^2$,
hence for $m_D >> 3T$  Eq. (\ref{sigma_md}) acts as an almost isotropic cross section and we should 
recover the results of the previous Section. On the other hand, we notice that the well known HTL estimate
of a gluon plasma viscosity \cite{ArnoldII} is valid only in the limit of $g=m_D/T<<1$, i.e. for very anisotropic 
cross section.

We have seen that the modified RTA gives the same result of the first order CE for isotropic cross section.
In order to perform the same analysis for the non-isotropic case we have to calculate the transport
cross section $\sigma_{tr}$:
\begin{equation}
\sigma_{tr} (s)=\int \, \frac{d\sigma}{dt} \, \sin^{2}\Theta \, dt = \sigma_{tot} \, h(a) 
\end{equation}
where $h(a)=4 a ( 1 + a ) \big[ (2 a + 1) ln(1 + 1/a) - 2 \big ]$ and $a=m_{D}^2/s$. For $m_{D}\to \infty$ the function $h(a) \to 2/3$ 
and  we recover the isotropic limit, $\sigma_{tr}=(2/3)\sigma_{tot}$, while for finite value of $m_D$ the function $h(a)<2/3$. 
One can generalize the concept of transport relaxation time $\tau^{-1}_{tr}=\langle \rho \, \sigma_{tr} \, v_{rel}\rangle$ that for 
isotropic cross section is equal to $(2/3)\langle \rho \, \sigma_{tot} \, v_{rel}\rangle$ writing

\begin{equation}
\tau^{-1}_{tr}=\langle \rho \, \sigma_{tr} \, v_{rel}\rangle= \rho \, \sigma_{tot} \, \langle v_{rel} \, h(a) \rangle
\end{equation}

The shear viscosity $\eta^*_{RTA}$ in the modified RTA is therefore given by 
\begin{eqnarray}
\eta^*_{RTA} = 0.8 \frac{1}{ \langle v_{rel} \, h(a) \rangle} \, \frac{T}{\sigma_{tot}}
\label{eta_anis_relax}
\end{eqnarray}
This is the same formula used in several transport calculation to fix the viscosity \cite{Abreu:2007kv,Ferini_PLB,Plumari_BARI},
but also the formula used to evaluate the viscosity in effective lagrangian approaches \cite{Sasaki:2008um}.
The thermal average,$\langle h(a) \, v_{rel} \rangle$, can be written more explicitely, see also \cite{koch}, as:
\begin{eqnarray}
\langle h(a) \, v_{rel} \rangle= \frac{8 z}{K^2_2(z)} \int_{1}^{\infty} dy\, y^2\,(y^2-1) \, h(2zy\, \overline a) \, K_1(2z y)=
f(z, \overline a )
\label{f_RTA}
\end{eqnarray}
where in the function $h$ we have re-written the argument as $a= 2zy\overline a$, with $\overline a= T/m_D$, just to
make explicit the integration over the $y$ variable. This leads to a compact formula for the viscosity in the RTA approximation:
\begin{eqnarray}
\eta^*_{RTA} = 0.8 \frac{1}{ f(z,\frac{T}{m_D})}\, \frac{T}{\sigma_{tot}}
\label{eta_RTA}
\end{eqnarray}
with the function $f$, defined by Eq.(\ref{f_RTA}), that essentially account for the ratio between the total and the
transport cross section and therefore it is a function of value smaller than $2/3$. Furthermore we notice that 
for massless particles it will be a function only of the $T/m_D$ ratio.

\begin{figure}
 \centering
 \includegraphics[scale=0.3 , keepaspectratio=true]{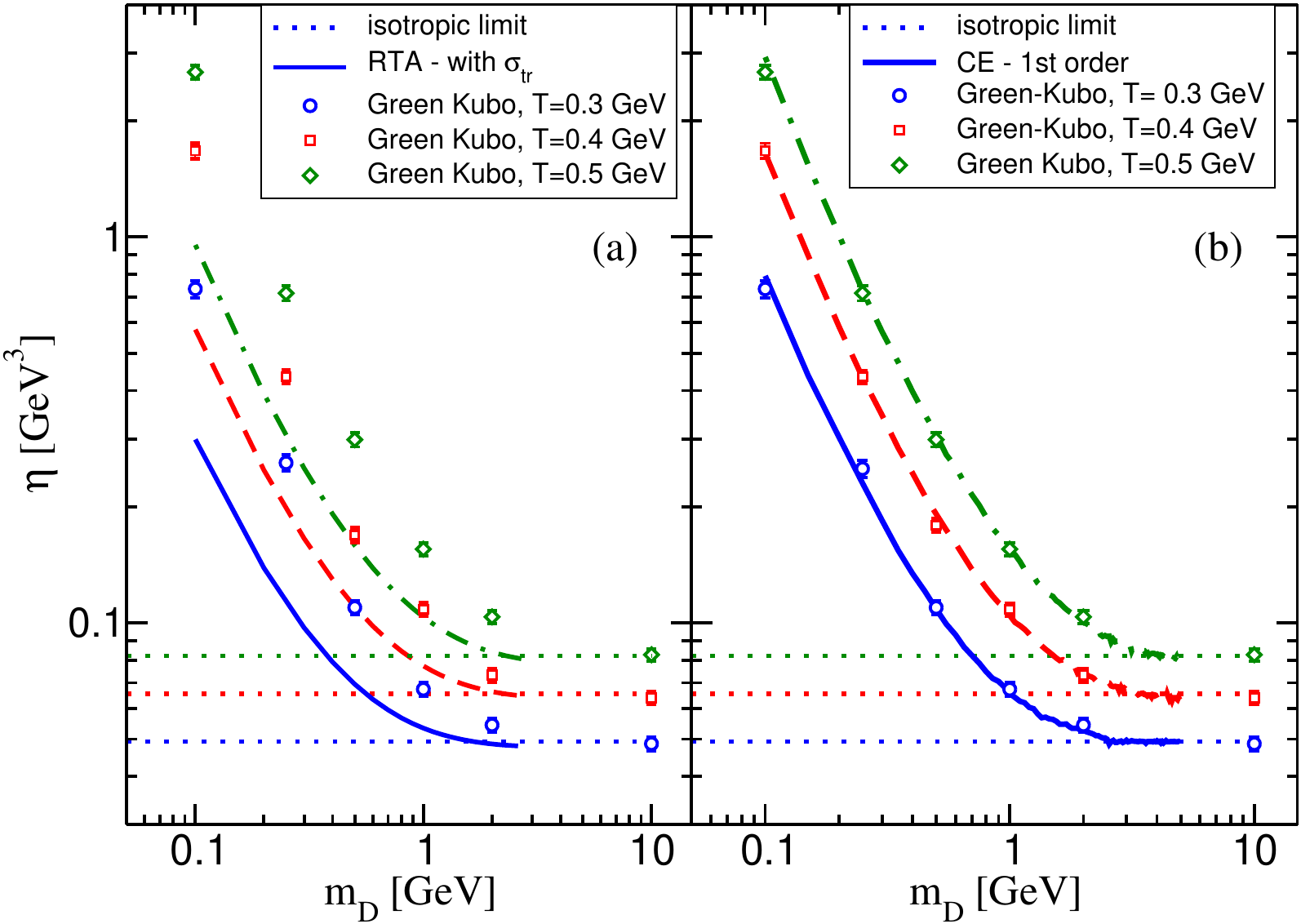}

 \caption{(Color online) Left: Shear viscosity $\eta$ as a function of the Debye mass $m_{D}$ for three different values of the temperature 
 $T=0.3 \, GeV$ (blue solid lines) and $T=0.4 \, GeV$ 
(red dashed lines) and $T=0.5 \, GeV$ (green dash-dot lines). The dotted line is the isotropic limit when $m_D \gg 1 \, GeV$, the solid line is the 
relaxation time approximation with $\tau^{-1}=\langle \rho \, \sigma_{tr} \, v_{rel} \rangle$ and 
dashed line is the Chapman-Enskog approximation at first order.
Right: The lines represent Chapman-Enskog approximation at the first order for the three different temperature 
($T=0.3, 0.4, 0.5\, \rm GeV$) while the open circles are the results obtained using the Green-Kubo relation.}
 \label{fig:eta_md}		
\end{figure}

In the CE approximation the situation is more complex, but for our case it is possible to write down the viscosity 
in a similar way obtaining, after some manipulation of Eq.(\ref{shear_I}), the following form:
 \begin{eqnarray}
 [\eta_s]^I_{CE} = 0.8 \,\frac{1}{g(z,\overline a)} \frac{T}{\sigma_{tot}}
\label{eta_CE_anis}
\end{eqnarray}
with 
\begin{eqnarray}
 g(z,\overline a)=\frac{32}{25} \frac{z}{K^2_3(z)} \int_{1}^{\infty} dy\, (y^2-1)^3 \, h(2zy\, \overline a) \, 
\left[  (z^2y^2+1/3)K_3(2z y) - zy K_2(2zy) \right] 
\label{g_CE}
\end{eqnarray}
it is clear that the function $g(z,T/m_D)$ is quite different from the thermal average of the transport cross section
as in the modified RTA, Eq.(\ref{f_RTA}). From the result shown in Fig.\ref{fig:eta_md} comparing the
lines on the left panel, evaluated from Eq. \ref{eta_RTA}, and the lines  on the right, evaluated from
Eq.(\ref{eta_CE_anis}), we see that the $g(z, T/m_D)$
is generally smaller that the thermal average of the transport cross section, see the comment below. 

%In physical terms it means
%that at fixed total cross section the increase in the shear viscosity is larger than what come from what come
%from  $\langle h(a) \, v_{rel} \rangle$ coming from the ratio of the transport to the total cross section.

In the left panel of Fig.(\ref{fig:eta_md}) it is shown the shear viscosity $\eta$ as a function of the Debye mass at fixed total 
cross section at $\sigma_{tot}=3 \,\rm mb$ and for three different temperatures $T=0.3 \, \rm GeV$, 
$T=0.4 \, \rm GeV$ and $T=0.5 \, \rm GeV$. 
The solid, dashed and dot-dashed lines are the behavior of $\eta$ in the modified RTA approximation, Eq.(\ref{eta_anis_relax}),
while the symbols are the result with the Green-Kubo formula. It is evident that there is a strong disagreement
between the two as soon as $m_D$ is such to move from the isotropic limit indicated to guide the eye
by dotted lines (which corresponds to the standard RTA).
Therefore, we see in general that even if the total cross section is kept constant, the anisotropy of the cross section
cause a strong enhancement of the viscosity $\eta$. However the increase is very strong and the difference
between transport and total cross section is not able to account fully for such an increase.
On the right panel of Fig. (\ref{fig:eta_md}) we compare the Green-Kubo results (symbols) with 
the prediction of CE at first order (solid, dashed and dot-dashed lines). In this case we find
a very good agreement between the two, hence the CE already at first order is able to account
for the correct value of $\eta$ even if the cross section is so forward-backward peaked to cause
an increase of about an order of magnitude respect to the same total cross section but isotropic.
The RTA approximation would severely underestimate the viscosity.
We can also see that for $m_D \sim 8-10 T $ the isotropic limit is recovered and both CE and RTA
coincide but this is essentially the limit discussed in the previous Section.
We note that the calculation have been performed down to quite low value of the screening mass,
$m_D= 0.1 \, \rm GeV$. This for $T= 0.5 \, \rm GeV$ would correspond to anisotropic cross section
that in the HTL approach corresponds to $g=m_D/T=0.2$.  Nonetheless within a precision
of about $5\%$ the first order CE is able to account for the correct value of $\eta$ even for such 
forward-backward peaked cross section. This result further validates the approach in Ref.\cite{ArnoldII}. 

In the introduction, we discussed the relevance that the knowledge of the analytical relation between the
shear viscosity $\eta$ and the microscopic and macroscopic quantities $T,\sigma(\theta),\rho,M$ can have
in building up a transport code that follow the evolution of a fluid at fixed $\eta/s$.
The idea is to fix the local cross section by reverting Eq.(\ref{eta_RTA}), i.e.
 $\sigma_{tot}=0.8\, T/ f(\overline a)\eta $.
First tentatives to develop such a transport approach were based on Eq.(\ref{eta_RTA}) \cite{Molnar_cascade,Ferini_PLB,Greco:2008fs,
Plumari:2010fg,Abreu:2007kv}, i.e. on the RTA
approximation. Our work shows that for realistic case where the screening mass $m_D \sim 0.3-1 GeV$
for the temperature range $T \sim 0.15-0.6 \rm GeV$ explored at RHIC and LHC energies,
Eq.(\ref{eta_CE_anis}) provides a sufficiently correct expression within a $5\%$.  Instead the use of 
Eq.(\ref{eta_RTA}) leads to underestimate the needed local cross section by about a factor
$1.5-2$, because as we have seen from our study $f(\overline a) > g(\overline a)$  except
for the isotropic cross section case where they are equal.

\section{Viscosity of a Gluon Plasma}

In this last part of the paper we want to apply the analysis performed for quite general case to
a realistic case of the shear viscosity to entropy density ratio for a gluon plasma.
Therefore we consider massless gluons in thermal equilibrium interacting via two-body collisions 
corresponding to the sum of $u-$ and $t-$ channels:
\begin{equation}\label{differential_cross_section}
\frac{d\sigma ^{gg\to gg}}{d q^2}=9\pi\alpha^2_s\frac{1}{(q^2+m^2_D)^2}.
\end{equation}
where $m_D$ is the Debye mass, $m_D=T\sqrt{4\pi \alpha_s}$ according to HTL calculations and 
\begin{equation}
\alpha_s(T)=\frac{4\pi}{11 \ln \left( \frac{2\pi T}{\Lambda} \right)^2}\, \,\,\,\, , \qquad \Lambda=200\,MeV
\end{equation}
is the pQCD running coupling constant.
A full pQCD calculation with HTL dressed propagator would include also the $s-$channel and all the interference terms,
however it has been shown that the $u-$ and $t-$ channels are the dominant ones \cite{ArnoldII}. On the other a full HTL calculation
as in Ref. \cite{ArnoldII} would require an energy dependent propagator with both longitudinal and transverse components, 
however our objective here it is not
to have the best evaluation of $\eta/s$, but to discuss the comparison between CE, RTA and the numerical
Green-Kubo method for a quite realistic case and also under the same condition of previous work based on the Green-Kubo method 
\cite{Fuini_3}.

The total cross section in this scheme is energy and temperature dependent:
\begin{equation}\label{total_cross_section}
\sigma_{tot}=\frac{9\pi\alpha^2_s}{\,m^2_D} \frac{s}{ s+m^2_D}.
\end{equation}

\begin{figure}
 \centering
 \includegraphics[scale=0.35, keepaspectratio=true]{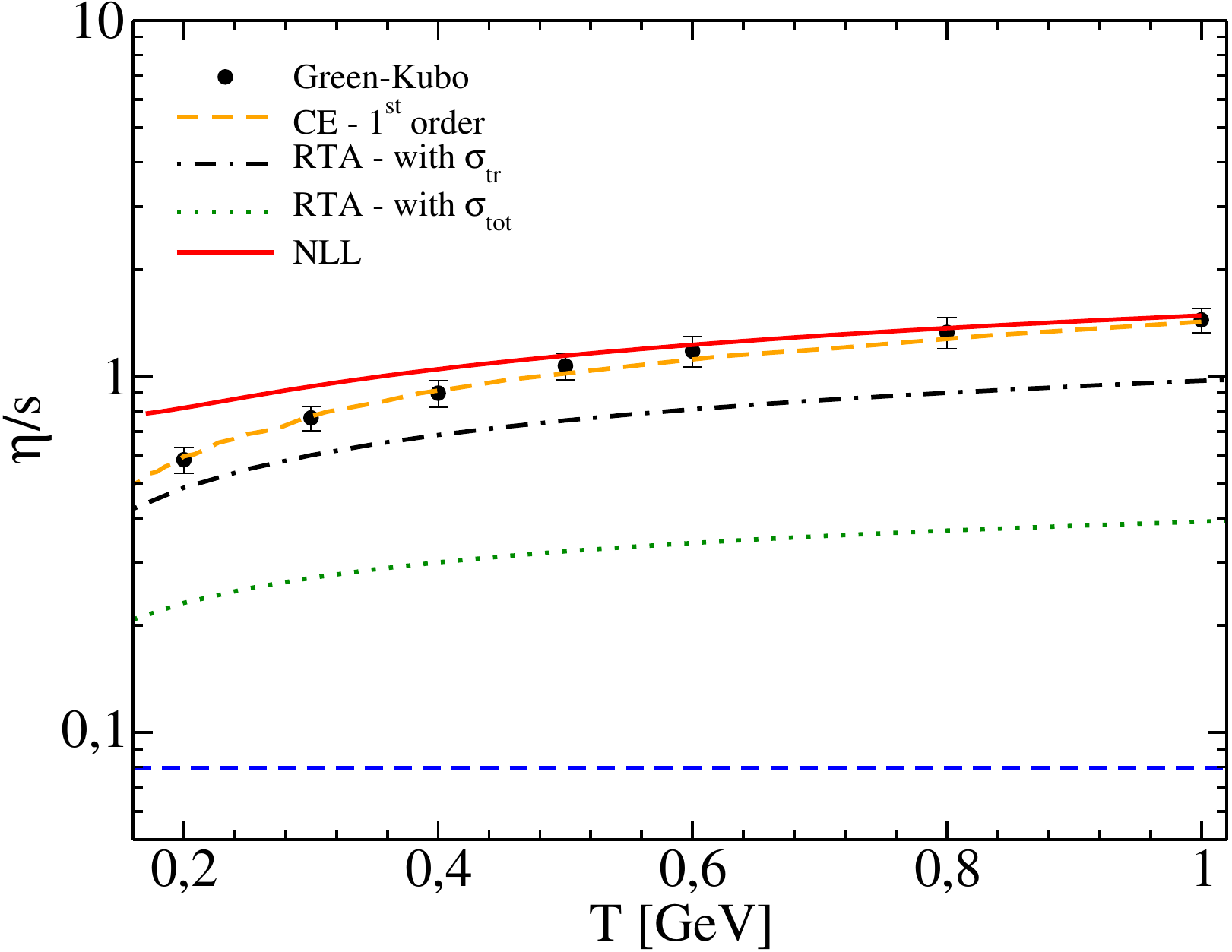}%{eta_su_s_temp_2.pdf}
 \caption{(Color online) Shear viscosity to entropy density ratio $\eta/s$ of a gluon plasma interacting through the differential cross section in Eq. (\ref{differential_cross_section}) as a function of temperature $T$: 
 black circles are the results obtained using Green-Kubo correlator, orange dashed line represents Chapman-Enskog first order approximation. 
The dot-dashed line is the prediction with the modified RTA approximation and the dotted lines is the standard RTA. Red line is the Next-to-Leading-Log order calculation from  Ref.\cite{ArnoldII}.
}
 \label{fig:etas_finale}		
\end{figure}

For this realistic case we have calculated the $\eta/s$ as a function of the temperature 
for different temperatures in the range $0.2\leq T \leq 1.0\,\rm GeV$ by mean of the numerical
Green-Kubo method. 
We notice that the entropy density $s/T^3$ is constant as for a free gas, see also par. 3.1.
The results shown by red symbols in Fig. \ref{fig:etas_finale} are compared to
the CE (orange dashed line), the modified RTA (black dot-dashed line) described above and  the simple RTA ( green dotted line)
with the relaxation time proportional to the total cross section and not to the transport one. We find again that the CE again is in excellent
agreement with the Green-Kubo result at the level of accuracy of $4\%$, while the modified RTA significantly underestimates 
the $\eta/s$ by about a $20\%$ at $T\sim 0.2 \,\rm GeV$. At increasing temperature the discrepancy tends to increase
up to about a $50\%$; 
we can understand this result on the base of the discussion in the previous section because the
$m^2_D/T^2 \propto \alpha_s(T)$ ratio becomes smaller at increasing temperature
 and therefore the cross section appear effectively less anisotropic.

We notice that for $T> 0.4 \,\rm GeV$ we predict a $\eta/s$  in good agreement with
the extrapolation of calculations in Ref.\cite{ArnoldII}, even if however we have not used the full HTL propagator.

A significant discrepancy is found respect to Ref.  \cite{Fuini_3}  especially for small  temperatures,
 while for temperatures above $T\geq 0.5\,GeV$ there is a tendency toward an agreement. 
 We have however used the same cross section 
in \cite{Fuini_3} hence in principle we should have a very similar $\eta/s(T)$. However
for $T < 0.5\,GeV$  they found an increasing of $\eta/s$, while our results have a monotonic logarithm 
increase as a function of temperature as one would expect from the running coupling $\alpha_s(T)$
and the T dependence of the screening mass. It appears difficult to justify even qualitatively
a non monotonic behavior of $\eta/s(T)$.
In fact from the RTA approximation one has parametrically $\eta/s \sim T/\sigma_{tot} T^3$ and
for the case of the cross section in Eq.(\ref{total_cross_section}) follows $\eta/s \sim 1 /\alpha_s(T)  \sim \,ln(T/\Lambda)$,
which is an increasing function of T at all temperatures. 
%We notice that this parametric dependence is however
%different from the one in ref.\cite{ArnoldII} and this is due to the fact that there an energy dependent self-energy
%for the propagator is considered which leads to a different relaxation time that can develop also a shallow
%minimum.

\section{Conclusions}

The recent developments in the physics of strong interaction at finite temperature T and baryon chemical potential $\mu_B$
is focusing many efforts on the calculation of transport coefficients and in particular of the shear viscosity.
There are several approximation schemes that relates the viscosity $\eta$ to the microscopic cross section.
The validity of such approximations, at least at the typical scales of quark-gluon plasma, was not yet investigated. 
We have developed
a method to solve numerically the Green-Kubo formula for the shear viscosity for the case of a relativistic Boltzmann gas.
Our objective was to compare the Chapman-Enskog approximation and the relaxation time approximation
 with the result from the Green-Kubo correlator that in principle should give the correct result.
We have performed such a comparison from a variety of physical case: isotropic and non-isotropic cross section
massless and massive particles at different temperatures.
Our work shows that the relaxation time approximation with a relaxation time $\tau \propto \sigma_{tot}$
always underestimate the shear viscosity even by more than a factor of 2-3. 
Moreover even the modified RTA with $\tau \propto \sigma_{tr} $ gives
a satisfying prediction only in the unrealistic case of massless particles and isotropic cross section. 
For realistic pQCD-like cross section one has on average a $30-40\%$ underestimate for $\eta$
respect to the Green-Kubo calculations.
Instead the Chapman-Enskog appears
to be a much better approximation scheme that already at first order have shown an agreement 
at the level of $4\%$ with the
numerical calculations of the Green-Kubo correlator for all the physical case considered including 
very forward peaked cross section and massive particles.

We note that the agreement of CE approximation at first order with the Green-Kubo method
also supplies a relatively simple analytical expression that can allow to
developed kinetic transport theory at fixed viscosity with very good precision.
Our work shows that the current used simple relation $\eta/s=0.8 T/\sigma_{tr}$
can lead to significantly underestimate, even by about a factor of two,
the $\eta/s$. Therefore previous works \cite{Ferini_PLB,Abreu:2007kv,Plumari_BARI} in this direction 
are only approximately valid.

In the next future it would be interesting to extend the study also to a collision integral including 
three-body collisions whose role in the equilibration dynamics in ultra-relativistic
heavy-ion collisions \cite{Xu:2004mz,Xu:2008av,Zhang:2010fx} and in the determination of the $\eta/s$ is of current 
interest.

Finally, it would be interesting to apply the method for the calculation of $\eta/s$ for a quark-gluon plasma,
this could allow also a more precise evaluation of this quantity in quasi-particle models where till now only
rough approximation for the relaxation time have been considered 
\cite{Sasaki:2008um,Bluhm:2010qf,Plumari_njl,Plumari_qpmodel}.

\section{Acknowledgements}
S. Plumari and A. Puglisi thank C. Wesp and  V. Greco  thanks D. Molnar for fruitful discussions. 
V.G. acknowledges the support by the ERC under the QGPDyn Grant.%

%\bibliographystyle{h-physrev3}
%\bibliography{comparisonreferences}

\end{document}